\documentstyle[preprint,aps,graphicx]{revtex}
\tightenlines
\begin{document}
\newcommand{\bref}[1]{eq.~(\ref{#1})}
\newcommand{\be}{\begin{equation}}
\newcommand{\en}{\end{equation}}
\newcommand{\bs}{$\backslash$}
\newcommand{\us}{$\_$}

\title{Interaction between a magnetic domain wall and a superconductor}
\author{L.E. Helseth, P.E. Goa, H. Hauglin, M. Baziljevich and T.H. Johansen}
\address{Department of Physics, University of Oslo, P.O. Box 1048 Blindern, N-0316 Oslo, Norway}

\maketitle
\begin{abstract}
The interaction between a magnetic thin film and a superconductor is studied.
In particular, the equilibrium width of a Bloch wall is estimated with and
without the superconductor. It is shown that the Bloch wall 
experiences a small shrinkage on cooling through the critical temperature of the
superconductor. Furthermore, the interaction between the Bloch wall and a single 
vortex is estimated.
\end{abstract}

\narrowtext

\newpage

\section{Introduction}
The interaction between superconductivity and magnetism has been studied
for several decades. Systems composed of alternating magnetic and
superconductive layers are of interest not only because they are model systems 
for the interplay of competing superconducting and magnetic order parameters,
but also because of numerous possible applications. Recently, the development 
of magnetic thin film technology has triggered new interest in this 
field.    
Of particular importance has been the possibility of examining the 
interaction between superconductivity and magnetism in high-temperature 
superconductors\cite{Bulaevskii1,Bulaevskii2,Santiago,Sonin1,Sonin2}.   

Bulaevskii $et$ $al.$ showed that magnetic domain structures in a magnetic film
in close contact with a superconducting film may enhance pinning of vortices,
since this gives an opportunity to pin the magnetic flux of the vortex rather
than its core\cite{Bulaevskii1}. It was suggested that the pinning of vortices
in superconductor/ferromagnetic multilayers can be 100 times greater than the
pinning by columnar defects. Later, this proposal has been partially verified, 
but only in the case of a bilayer structure\cite{Santiago}.

Another interesting proposal is that of Sonin, who suggested that the
magnetostatic field from a domain wall may create a weak link at which single 
vortices could be localized\cite{Sonin1}. Then, by moving the domain wall, 
one should also be able to move the weak link as well. 

Evidently, many interesting applications could be developed if such interactions 
are better understood. In the present work we examine the interaction between a 
magnetic domain wall and a superconductor. First we investigate the interaction 
between a thin magnetic film and a superconducting substrate, and 
estimate the equilibrium width of a Bloch wall in the film. It will be shown that due to flux 
repulsion, the domain wall experiences a small shrinkage on cooling through the 
critical temperature of the superconductor. This could be exploited in 
magnetooptic waveguide systems. In such systems it is possible to match the
interacting mode's phases using the spatial periodicity of a sequence of Bloch
walls\cite{Kotov}. Altering the width of these domain walls simply by tuning the
temperature could be an effective way to change the light propagation in the
waveguide. 

We also study the interaction between the domain wall and a single vortex in a
type II superconductor. This is of interest both in fundamental and applied
physics. If we build further on the idea of Sonin, it should be possible to 
create a memory device based on active control of generation and annihilation 
of vortices by means of one or more domain walls. In recent years 
superconducting circuits based on single-flux-quantum pulses have been shown to provide a new family of digital 
electronics with ultra-high speed and very low power-dissipitation. At clock 
rates exceeding 10 GHz and operation speed of many hundred GHz, these devices 
can in the future outrun any semiconductor device\cite{Wang}. 
Using domain walls as active 'vortex-gates' we may add an additional degree 
of freedom in these devices. It is known that bismuth-substituted ferrite 
garnet films with in-plane magnetization have domain walls with very low 
coercivity that can be moved  without ambiguity at frequencies up to several 
GHz\cite{Kotov}. Furthermore, in such materials Bloch walls are easily formed by external
magnetic fields or stress patterns, and these could be manipulated in 
numerous ways suitable for a memory device.

\section{Equilibrium wall width}
Consider a magnetic film of thickness $d$ with two domains of opposite in-plane 
magnetization. The domains are separated by a 180 degree Bloch wall of
width $w$ and length $L$. The magnetic film is placed in contact with a type I
superconductor. We assume that the superconductor has zero penetration depth  
so that an image of the Bloch wall is formed inside the superconductor as shown
in Fig. \ref{f1}. Here we want to estimate the equilibrium 
wall width with and without the superconductor. To this end, we use a linear 
wall model,
\begin{equation}
\theta =\frac{\pi x}{w}, \,\,\,   -w/2<x<w/2   \,\,\, ,
\end{equation}
where $\theta$ is the angle between the magnetization vector and z-axis. 
Contributions to the total wall energy comes from the
exchange interaction, the crystalline anisotropy, the magnetostatic energy
and magnetoelastic effects. Here we neglect the magnetoelastic energy, which is
justifiable when the wall width is small. Also, if the substrate on which the
magnetic film is deposited is thick, a large portion of the stress is 
dissipated in the substrate as well (note that this substrate is not 
necessarily the superconductor, but could be some other material on which the
magnetic film is deposited).
  
For a Bloch wall the anisotropy energy per unit length of wall is given by,
\begin{equation}
E_{u}=wd\frac{1}{\pi}\int_{-\pi /2}^{\pi /2} K_{u}sin^{2} \theta d\theta
=\frac{1}{2} wdK_{u}\,\,\, , 
\end{equation}
where $K_{u}$ is the uniaxial anisotropy constant.

The exchange energy per unit length of wall is expressed by,
\begin{equation}
E_{ex}=wd A\left( \frac{\partial \theta}{\partial x} \right) ^{2} 
=\pi ^{2}A\frac{d}{w} \,\,\, ,
\end{equation}
where $A$ is the effective exchange constant.

The magnetostatic energy of a Bloch wall can be found by approximating the wall
with a homogenously magnetized elliptic cylinder\cite{Cullity},
\begin{equation}
E_{m}=\frac{1}{2}\mu _{0}\frac{w^{2} d}{w+d} M_{s}^{2} \,\,\, ,
\label{mag}
\end{equation}
where $\mu _{0}$ is the permeability of vacuum, and $M_{s}$ is the saturation
magnetisation in the magnetic material. Eq. \ref{mag} is a reasonable
approximation for materials with low permeability, and has been used to model
the domain wall behaviour in ferrite garnet films (see e.g. ref.\cite{Kryder}
and references therein).

In the presence of the superconducting substrate the magnetic surface charge 
at $z=0$ is at most doubled, which means that the energy
density cannot increase by more than four times. In the limit $d\gg w$ the energy density at 
$z=d$ is not altered. If we now assume that the area occupied by the magnetic field is not decreased, then the average
energy in presence of the superconductor is $(4+1)/2=2.5$ times that without the
superconductor. This is an upper estimate of the increase in energy, 
since the area will change upon introduction of the superconductor, and the
energy density is lower than that assumed here. To date, nobody has performed 
an accurate analysis of the magnetostatic energy resulting from the influence 
of a superconducting substrate. However, it is reasonable to expect that the
magnetostatic energy has a similar functional dependence of film thickness and 
wall width as in eq. \ref{mag}, if we assume that only the width of the 
domain wall changes upon introduction of the superconductor. Therefore we will 
characterize the increase in energy by a factor $\gamma$ 
\begin{equation}
E_{m}=\frac{\gamma}{2}\mu _{0}\frac{ w^{2}d}{w+d} M_{s} ^{2} \,\,\, ,
\end{equation}
where $\gamma =2$ with and $\gamma =1$ without the superconductor. We strongly
emphasize that $\gamma =2$ is only a reasonable guess made in order to estimate an
upper bound for the superconductor's influence on the wall width, and that a
complete micromagnetic analysis is required to obtain a more accurate answer. 
Such an analysis should take into account the finite penetration depth and the fact
that the magnetization in the Bloch wall changes in a continuos manner. 
   
To find the equilibrium wall width, we must minimize the total energy according
to 
\begin{equation}
\frac{\partial E}{\partial w} =0 \,\,\, .
\end{equation}
Here we will only consider the limit $d\gg w$,
\begin{equation}
\gamma \mu _{0} M_{s}^{2} w^{3} + \frac{1}{2} dK_{u}w^{2}  - \pi ^{2}Ad = 0\,\,\, ,
\label{ab}
\end{equation}
which can easily be solved numerically.

Increasing the magnetization increases the effect of the superconducting 
substrate as well. It is seen that when the anisotropy constant
can be neglected, the equilibrium wall width becomes,
\begin{equation}
w= \left( \frac{\pi^{2} Ad}{\gamma \mu _{0} M_{s}^{2}} \right) ^{1/3}\,\,\, ,
\end{equation}
and the wall width decreases at most by $2^{1/3} \approx 1.3$ by crossing the 
critical temperature of the superconductor. 

Sonin analyzed a periodic array of 
domains with magnetizations perpendicular to the film, and found that in the 
limit $K_{u} \gg 1/2\mu_{0} M_{s} ^{2}$ the domain width decreases at most by 
$\sqrt{1.5}$\cite{Sonin2}. In our case the change is probably smaller, since 
the contribution due to the uniaxial anisotropy is often 
comparable to that from the magnetization. However, when the 
magnetostatic energy can be neglected, the wall width is given by
\begin{equation}
w=\pi\sqrt{\frac{2A}{K_{u}}} \,\,\,  ,
\end{equation}
and the superconductor has no influence.

As an example we calculate the equilibrium wall width in the case of 
a ferrite garnet film of composition $Lu_{3-x}Bi_{x}Fe_{5-z}Ga_{z}O_{12}$. 
In these films it is easy to obtain single Bloch walls of the kind discussed 
here. Reasonable material parameters are $A \sim 2\cdot 10^{-11}$ $J/m$ and 
$K_{u} \sim 10^{3}$ $J/m^{3}$. Fig. \ref{f2} shows the wall width as a function 
of the magnetization $M_{s}$ with ($\gamma =2$) and without ($\gamma =1$) the 
superconducting substrate. It is seen that the wall 
width decreases with increasing magnetization. Note also that the difference
between $\gamma =1$ and $\gamma =2$ is around 20 $\%$. In a magnetooptic 
waveguide a 20 $\%$ change in the wall width is probably enough to alter the 
light propagation substantially. A larger difference can be obtained by 
reducing the anisotropy constant $K_{u}$. In 
$Lu_{3-x}Bi_{x}Fe_{5-z}Ga_{z}O_{12}$ this is often done by reducing the
Bi - content.   

\section{Interaction between a domain wall and a single vortex} 
Consider a straight vortex located a distance $a$ from the Bloch wall, see Fig.
\ref{f3}. Due to the magnetic field from the Bloch wall, there will be 
an interaction between the two. We assume that the magnetic film is so thick 
that the magnetic poles at $z=d$ and $z=-d$ do not 'feel' the field from the 
vortex, and the pole strength at $z=0$ is now two times that of the domain wall 
alone (if the distance $a$ is large and the penetration depth is small). 
To find the interaction between the vortex and the domain wall, one 
should in general solve the London-Maxwell equations, including the contributions from 
supercurrents. However, here we estimate only the purely magnetostatic
interaction, which means that the interaction energy can be calculated 
considering only the magnetostatic forces between a magnetic monopole and 
a magnetic surface charge, using the following integral
\begin{equation}  
E_{int} =
\mu _{0} \int_{S}^{} \phi \mbox{\boldmath $M$} \mbox{\boldmath
$\cdot$} \mbox{\boldmath $dS$}\,\,\,   .
\end{equation}

It has been found that the field from a vortex is similar to that from a 
magnetic monopole located a distance $z_{0}=-1.27\lambda$ ($\lambda$ is the 
penetration depth) below the superconductor surface\cite{Carneiro}. 
In this approximation the scalar potential can be written as,
\begin{equation}
\phi =\frac{\Phi_{0}}{2\pi \mu _{0}} \frac{1}{\sqrt{(x-a)^{2} + y^{2}
+(z-z_{0})^{2}}} \,\,\,  ,
\label{ac}
\end{equation}
where $\Phi _{0}$ is the flux quantum.
Note that eq. (\ref{ac}) assumes that the medium above the superconductor is
isotropic with permeability $\mu _{0}$. An accurate calculation should take 
into account the anisotropy of the magnetic film. However, here we will
neglect the fact that the magnetic film alters the field from the vortex, in order to
get a simple estimate of the interaction energy. Then the x-component of the 
force acting on the vortex is given by
\begin{eqnarray}
F_{x} &=&  -\frac{\Phi _{0} M_{s} }{\pi } \nonumber \\
& & \times ln \left| \frac{ \left[ \frac{L}{2} +\sqrt{(\frac{L}{2} )^{2}
+(\frac{w}{2} -a)^{2} +(z-z_{0})^{2} } \right] \left[ -\frac{L}{2} +\sqrt{(\frac{L}{2} )^{2}
+(\frac{w}{2} +a)^{2} +(z-z_{0})^{2} } \right]}{\left[ \frac{L}{2} +\sqrt{(\frac{L}{2} )^{2}
+(\frac{w}{2} +a)^{2} +(z-z_{0})^{2} } \right] \left[ -\frac{L}{2} +\sqrt{(\frac{L}{2} )^{2}
+(\frac{w}{2} -a)^{2} +(z-z_{0})^{2} } \right]} \right| \,  .
\label{force}
\end{eqnarray}
We have assumed that the magnetic charge is $\sigma =-2M_{s}$, which is strictly
valid for zero penetration depth and z=0. In the case that the magnetic wall is
moved away from the superconductor ($z \neq 0$), the magnetic charge changes, 
and eq. \ref{force} must be regarded as a rough approximation. It must 
also be pointed out that the vortex is a normal state region, and is therefore 
expected to change the magnetic charge when it is near the Bloch wall. Thus the 
expression for $F_{x}$ should be regarded as an upper limit of the force between the vortex 
and the Bloch wall, but should have the correct order of magnitude. 

When L is much larger than $\lambda$, $w$ and $a$, then $F_{x}$ is almost 
independent of L. Note that the interaction strength can be tuned by changing 
the magnetization, which could be useful in a potential memory device. To visualize the strength of the 
interaction for different magnetizations, Fig. \ref{f4} shows the force as a function of 
distance when $A = 2\cdot 10^{-11}$ $J/m$, $K_{u} = 10^{3}$ $J/m^{3}$ , 
$d=5$ $\mu m$, $\lambda =100$ nm, $L=100$ $\mu m$, z=0 nm, and 
eq. (\ref{ab}) was used to calculate the wall thickness $w$. The solid line 
correspond to $M_{s} =120$ $kA/m$ and the dashed
line to $M_{s} =70$ $kA/m$. The figure shows 
that by decreasing the magnetization, the force decreases as well. 
As expected from Fig. \ref{f3}, the vortex is attracted towards the domain wall from both
sides of the domain wall, and can be captured if it comes close enough. 
Note also that the vortex is repelled if the polarity of the Bloch wall is 
reversed. We see that the force $F_{x}$ is rather small. Thus only if the 
pinning strength is small enough, the Bloch wall can be used to move the 
vortex. To develop a memory device as discussed in the introduction, one 
needs carefully designed high-temperature superconductors with low pinning 
strength.

\section{Conclusion}
We have studied the interaction between a magnetic thin film and a 
superconductor. In particular, the equilibrium width of a Bloch wall is 
estimated with and without a superconducting substrate. It is shown that the 
Bloch wall experiences a $20$ $\%$ shrinkage on cooling through the critical 
temperature of the superconductor. 
Furthermore, the interaction between the Bloch wall and a single vortex is 
estimated, and it is found that the domain wall is able to trap the vortex if
the vortex comes close enough.

\section{Acknowledgements}  
This study was supported by the Norwegian Research Council.

\begin{figure}
\caption{A magnetic thin film with two in-plane magnetized domains placed on 
top of a superconducting substrate.
\label{f1}}
\end{figure}

\begin{figure}
\caption{The equilibrium wall width as a function of magnetization with
(dashed line) and without (solid line) the superconducting substrate. 
We have assumed that $A \sim 2\cdot 10^{-11}$ $J/m$, 
$K_{u} \sim 10^{3}$ $J/m^{3}$ and $d=5$ $\mu m$.
\label{f2}}
\end{figure}

\begin{figure}
\caption{The basic geometry for a Bloch wall located a distance $a$ from a 
single vortex.
\label{f3}}
\end{figure}

\begin{figure}
\caption{The force $F_{x}$ as a function of distance $a$ from the vortex when 
$M_{s} =120$ $kA/m$ (solid line) and $M_{s} =70$ $kA/m$ (dashed line) .  
\label{f4}}
\end{figure}

\newpage
\centerline{\includegraphics[width=14cm]{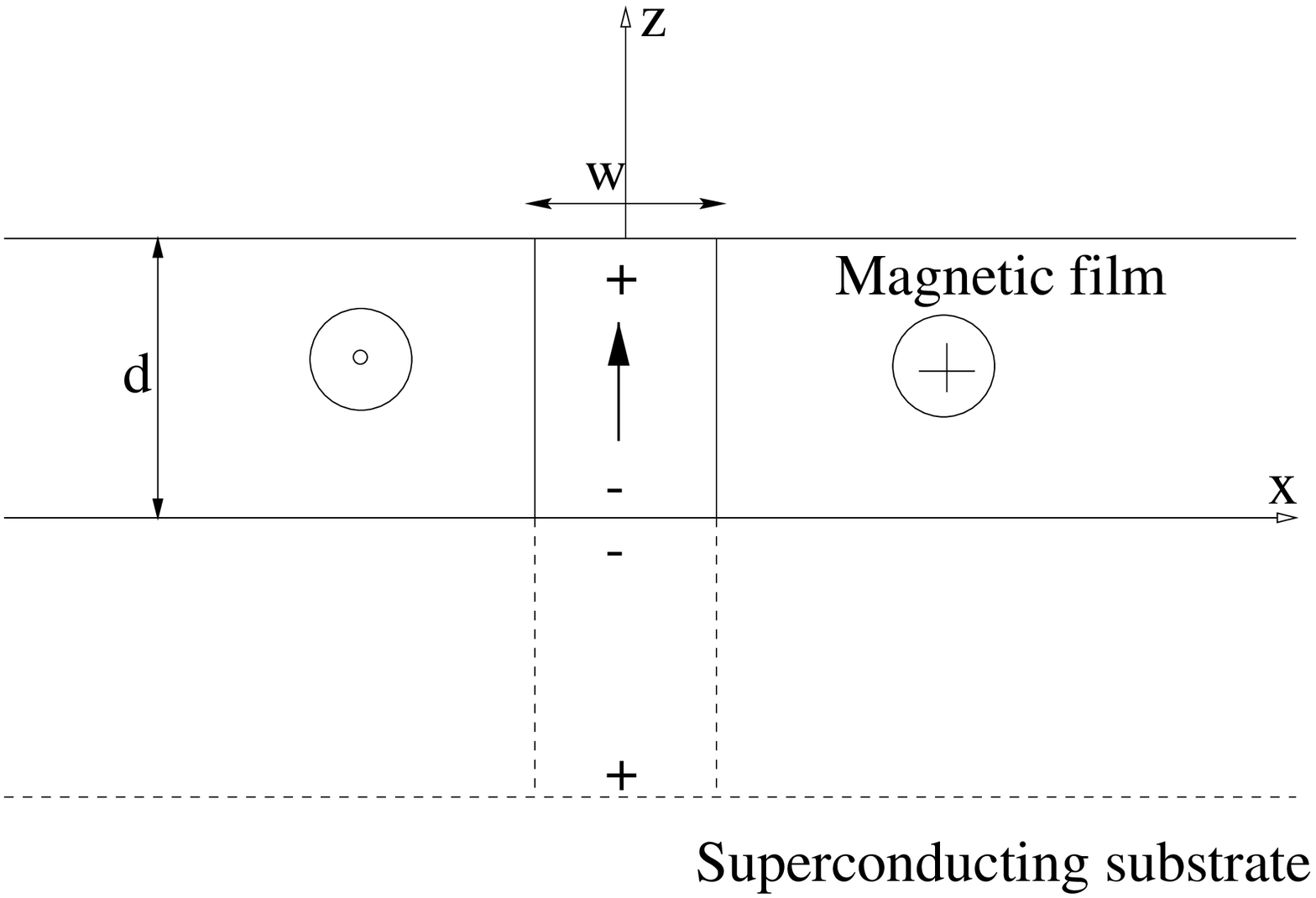}}
\vspace{2cm}
\centerline{Figure~\ref{f1}}

\newpage
\centerline{\includegraphics[width=14cm]{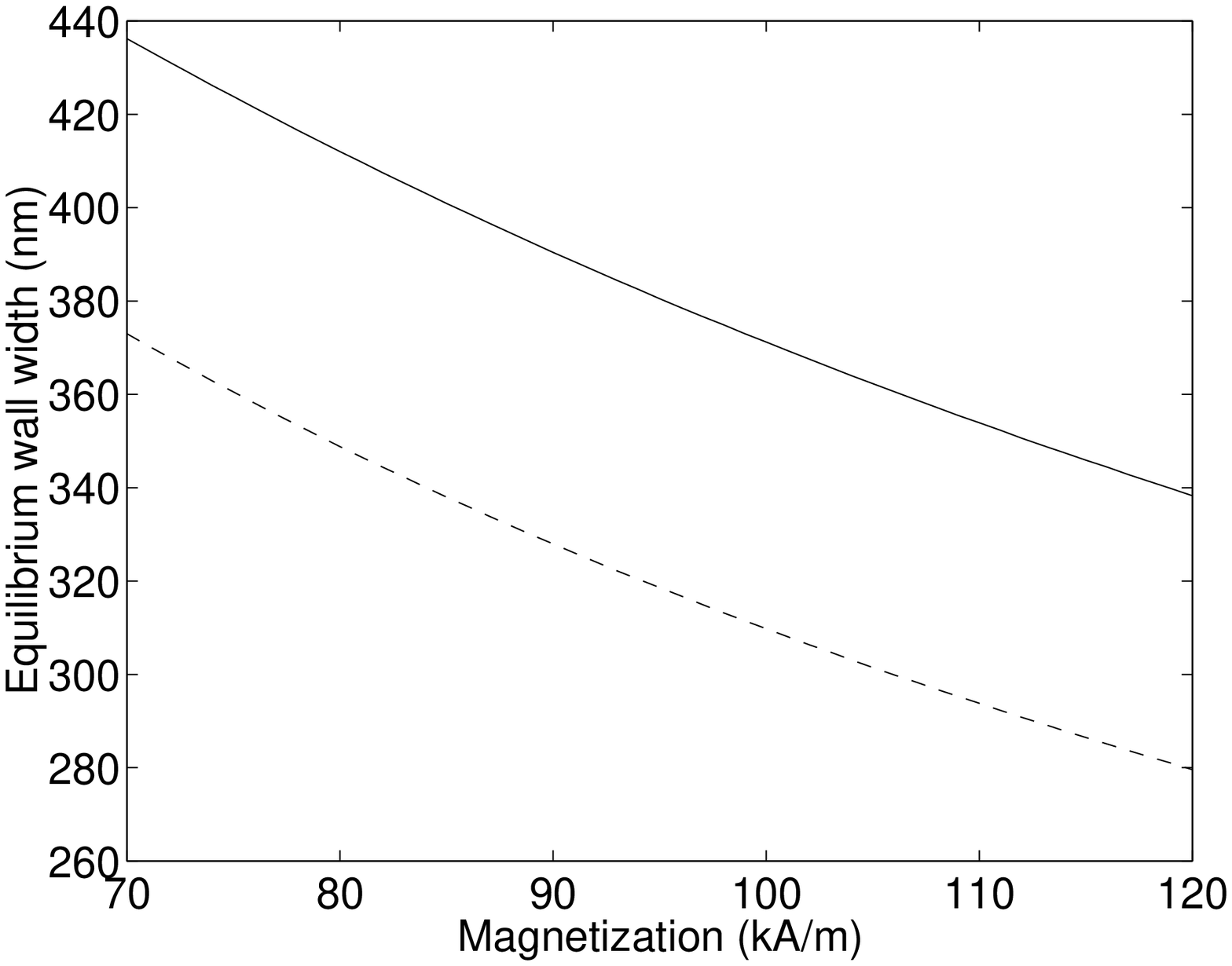}}
\vspace{2cm}
\centerline{Figure~\ref{f2}}

\newpage
\centerline{\includegraphics[width=14cm]{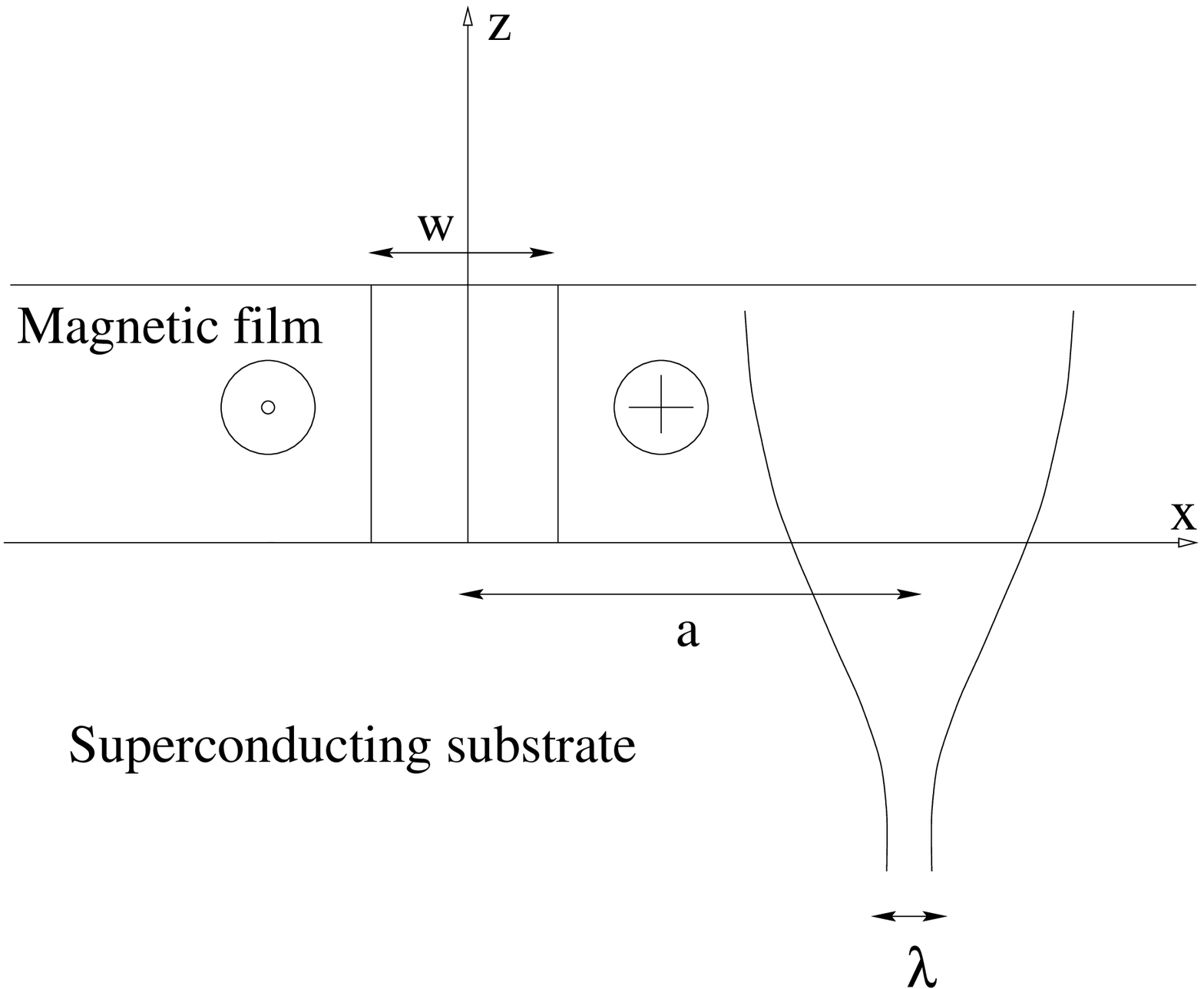}}
\vspace{2cm}
\centerline{Figure~\ref{f3}}

\newpage
\centerline{\includegraphics[width=14cm]{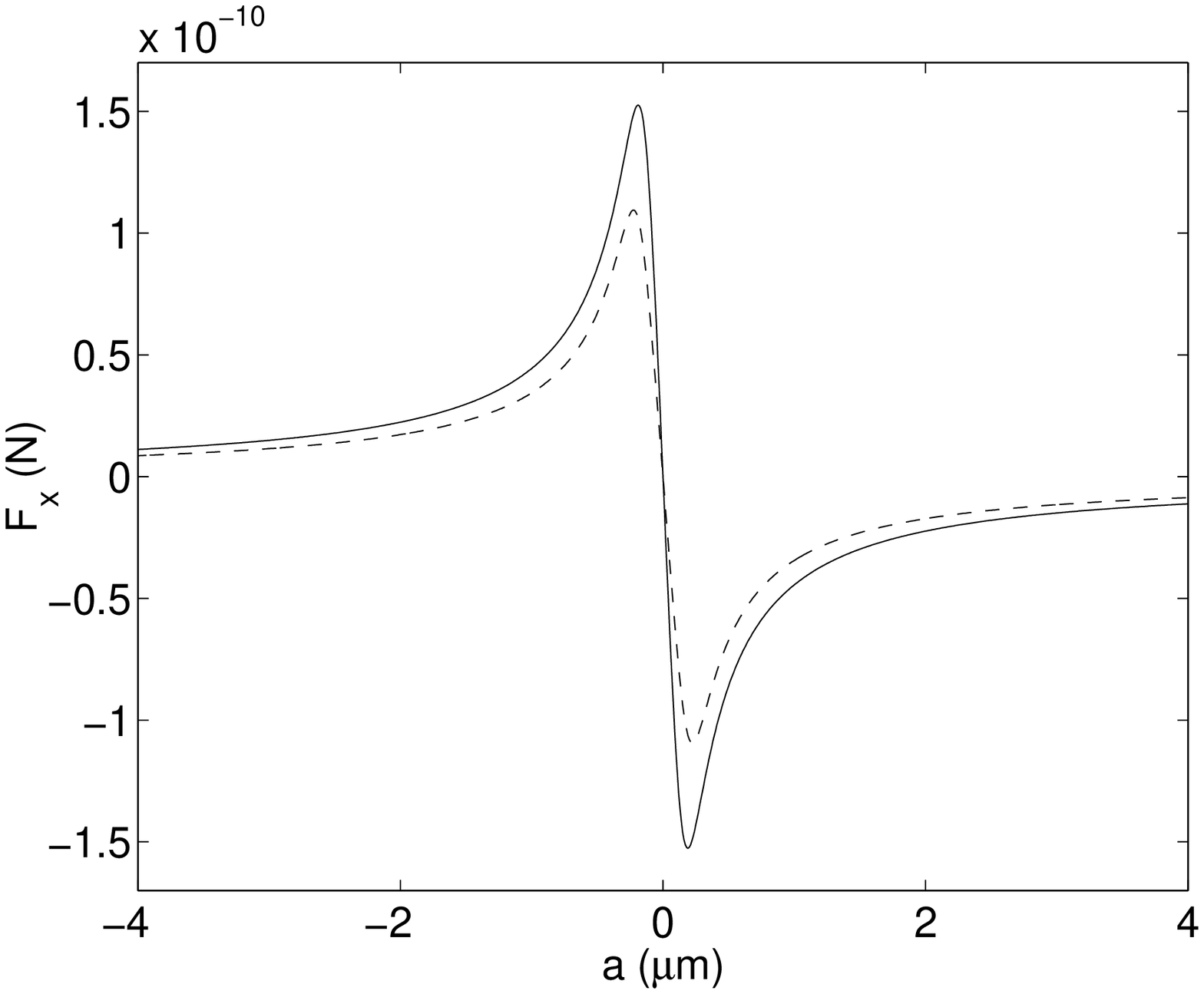}}
\vspace{2cm}
\centerline{Figure~\ref{f4}}

\end{document}